\documentclass{epsconf}
\usepackage{graphicx}
\usepackage{wrapfig}
\usepackage{amsmath,amssymb}

\newcommand{\reff}[1]{(\ref{#1})}
\newcommand{\eref}[1]{Eq.\reff{#1}}
\newcommand{\erefs}[1]{Eqs.\reff{#1}}
\newcommand{\figref}[1]{Fig.\ref{#1}}
\newcommand{\p}{\partial}

\newcommand{\omp}{\omega_p}
\newcommand{\citer}[1]{Ref.\cite{#1}}
\newcommand{\citers}[1]{Refs.\cite{#1}}
\usepackage{color}

\title{Resonance overlap and non-linear velocity spread \\ in Hamiltonian beam-plasma systems}
\author{\underline{N. Carlevaro}$^{1,2}$, G. Montani$^{1,3}$, F. Zonca$^{1}$}
\institute{$^1$ENEA, Fusion and Nuclear Safety Dep., C.R. Frascati, Via E. Fermi 45, 00044 Frascati, Italy.\\
           $^2$LTCalcoli Srl, Via Bergamo 60, 23807 Merate (LC), Italy.\\
           $^3$Physics Dep., ``Sapienza'' University of Rome, P.le Aldo Moro 5, 00185 Roma, Italy.}

\begin{document}
\maketitle

\paragraph{Abstract} We analyze some specific features of the beam-plasma instability. In particular, non-perturbative effects  in the dispersion relation are studied when the standard perturbative inverse Landau damping treatment breaks down. We also elucidate how only the global distortion of the profile rather than the clump width is truly predictive of resonance overlap at saturation.

\paragraph{Hamiltonian description of the beam-plasma interaction}
The beam-plasma system consists of a fast electron beam injected into a 1D plasma (taken as a periodic slab of length $L$), which is treated as a cold linear dielectric medium (having dielectric function $\epsilon=1-\omp^2/\omega^2$, $\omp$ being the plasma frequency) supporting longitudinal electrostatic Langmuir waves. The plasma density $n_p$ is assumed much greater than that of the beam $n_B$. The Hamiltonian formulation described in \citers{CFMZJPP,ncentropy} (and refs. therein) is here adopted and the broad beam is discretized in $n\gg1$ cold beams self-consistently evolving with $m$ modes at $\omega_j\simeq\omp$ for $j=1,\,...,\,m$ ($\epsilon\simeq0$). 
Resonant excitation occurs for mode wave numbers $k_j=\omp/v_{rj}$, where $v_{rj}$ is the initial velocity of a considered cold beam. The evolutive equations write \footnote{The particle positions along the $x$ direction are labeled by $x_i$, with $i=1,\,...\,N$ ($N$ being the total particle number): these are scaled as $\bar{x}_i=x_i(2\pi/L)$. The Langmuir scalar potential $\varphi(x,t)$ is expressed using the Fourier components $\varphi_j(k_j,t)$. We use the normalization: $\eta=n_B/n_p$, $\tau=t\omp$, $u_i=\bar{x}_i'=v_i(2\pi/L)/\omp$, $\ell_j=k_j(2\pi/L)^{-1}$, $\phi_j=(2\pi/L)^2 e\varphi_j/m\omp^2$, $\bar{\phi}_j=\phi_j e^{-i\tau}$. The prime indicates derivative with respect $\tau$ and barred frequencies and growth rates are $\bar{\omega}=\omega/\omp$ and $\bar{\gamma}=\gamma/\omp$, respectively.}
\begin{equation}\label{mainsys1}
\bar{x}_i'=u_i \;,\qquad
u_i'=\sum_{j=1}^{m}\big(i\,\ell_j\;\bar{\phi}_j\;e^{i\ell_j\bar{x}_{i}}+c.c.\big)\;,\qquad
\bar{\phi}_j'=-i\bar{\phi}_j+\frac{i\eta}{2\ell_j^2 N}\sum_{i=1}^{N} e^{-i\ell_j\bar{x}_{i}}\;,
\end{equation}
where resonance conditions rewrite $\ell_j u_{rj}=1$, with $\ell_j$ defined as integer. We set the initial warm beam distribution function in velocity space as $F_0(u)=0.5\;\textrm{Erfc}[a-b\,u]$  (with $a\simeq6.8$ and $b\simeq4537$). In the non-linear simulations of \erefs{mainsys1} (Runge-Kutta (fourth order) algorithm), we initialize $n=400$ cold beams having particle numbers distributed with $F_0$ (for a total $N=10^{6}$ particles) and equispaced velocities. Initial conditions for $\bar{x}_i$ are given uniformly in $[0,\,2\pi]$ for each cold beam, while modes are initialized as $\mathcal{O}(10^{-14})$ to ensure initial linear response.

\paragraph{Linear non-perturbative effects} The linear dispersion relation of the system writes
\begin{align}\label{disrel}
2(\bar{\omega}_0+i\bar{\gamma}_L-1)-\frac{\eta u_r}{M}
\int_{-\infty}^{+\infty}\!\!\!\!\!\!\!du\frac{\p_u F_0}{u/u_r-\bar{\omega}_0-i\bar{\gamma}_L}=0\;.
\end{align}
Here, the dielectric is expanded near $\omega\simeq\omp$ (according to motion equations), $M=\int_{-\infty}^{+\infty} du F_0$, and we used $\bar{\omega}=\bar{\omega}_0+i\bar{\gamma}_L$, where $\bar{\omega}_0$ denotes the real frequency shift and $\bar{\gamma}_L$ the growth rate of the considered mode. \eref{disrel} is numerically integrated (fixing $\eta$) for $u_r\in[0.0011,\,0.0019]$. The growth rates result to increase approaching the inflection point of $F_0(u)$, where the drive of the inverse Landau damping $\p_u F_0$ is maximum. Meanwhile, $\bar{\omega}_0$ starts to deviate from unity on the flat profile ($\p_u F_0\simeq0$): a finite growth rate (necessarily associated to a real frequency shift with respect to $\omp$) marks the break-down of the perturbative Landau damping expression.

The non-perturbative and non-local character of the dispersion relation can be enlightened by comparing the solution of \eref{disrel} with analytical and semi analytical treatments. In particular, linearizing \eref{disrel} (thus, by assuming $\bar{\omega}_0=1$), one can get the well know expression\vspace{-0.55cm}
\begin{wrapfigure}{r}{0.4\textwidth}\vspace{-0.2cm}
\includegraphics[width=0.4\textwidth]{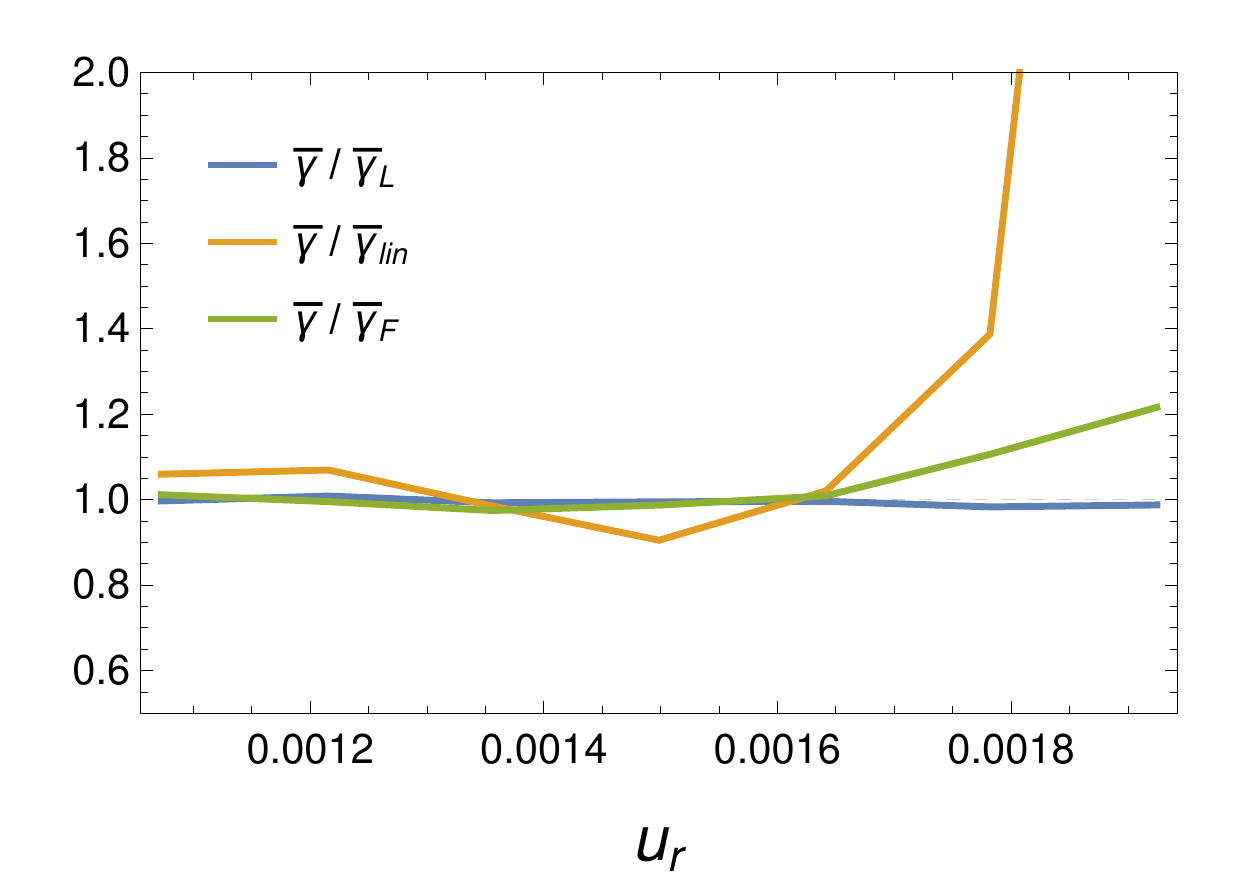}\vspace{-0.5cm}
\caption{\it \small Ratios between $\bar{\gamma}$ (simulation fit) and $\bar{\gamma}_L$, $\bar{\gamma}_{lin}$ and $\bar{\gamma}_F$ from \erefs{disrel}, \reff{drlin} and \reff{drfad}, as function of $u_r$ ($\eta=0.0007$).
\label{fig_gamma}}
\end{wrapfigure}\vspace{-0.cm}
\begin{align}\label{drlin}
\bar{\gamma}_{lin}-\pi\eta u_r^2\p_u f_B\big|_{u_r}/2 M=0\;.
\end{align}
Another approach consists in the integration of \eref{disrel} in terms of the Faddeeva function $w(x)$ and residue contributions \cite{malik16}:
\begin{align}\label{drfad}
&2(\bar{\omega}_{0F}+i\bar{\gamma}_F-1)+\nonumber\\
&\qquad-\eta A u_r\;i\pi(2e^{-X}-w(X))/M=0\;,
\end{align}
whose roots can be numerically determined (here $X=BC-BD(\bar{\omega}_{0F}+i\bar{\gamma}_F)u_r$, with $A=2559.88$, $B=5.89$, $C=1.15$ and $D=770.0$).

The ratios between $\bar{\gamma}$ fitted from the single mode simulations of \reff{mainsys1} and $\bar{\gamma}_L$, $\bar{\gamma}_{lin}$ and $\bar{\gamma}_F$ from \erefs{disrel}, \reff{drlin} and \reff{drfad}, are plotted in \figref{fig_gamma} vs. the resonant velocity ($\eta=0.0007$). The three methods correspond to different approximations of \erefs{mainsys1}, and the degree of accuracy of the specific solutions is evident from the plot. This indicates the non-perturbative character of the dispersion relation. Moreover, the integral over the whole distribution function and the non-linear frequency dependence make the estimation of the growth rate intrinsically non-local.

\begin{figure}[ht]
\includegraphics[width=0.317\textwidth]{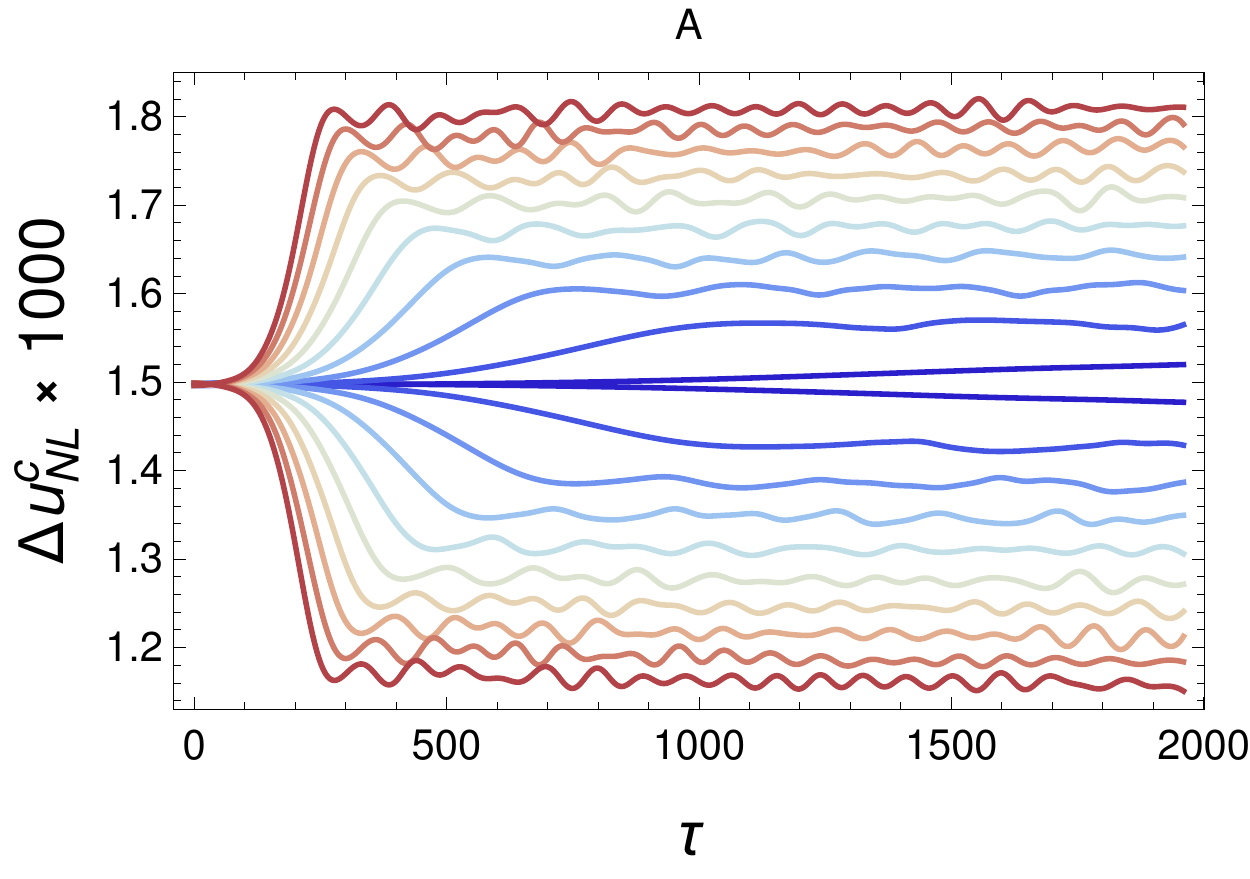}
\includegraphics[width=0.325\textwidth]{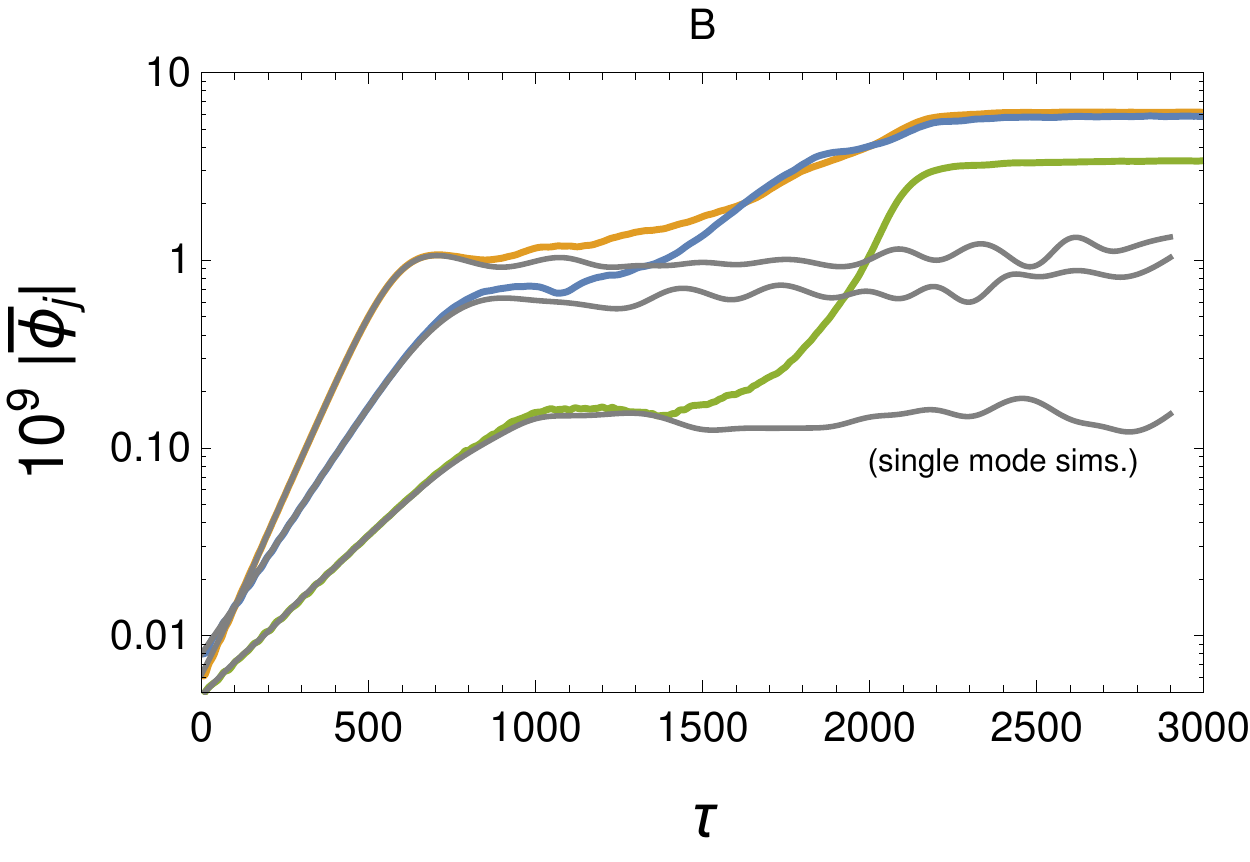}
\includegraphics[width=0.31\textwidth]{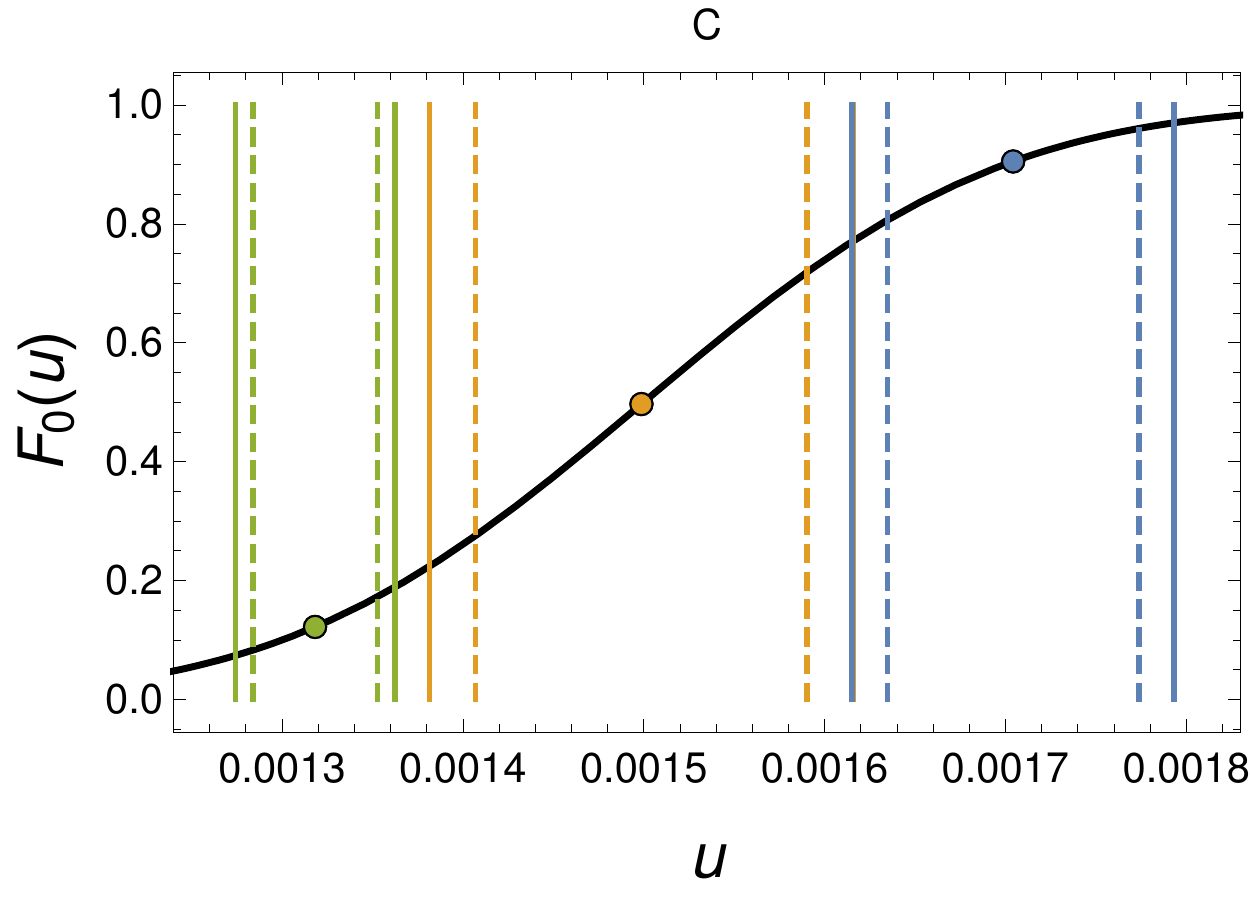}\vspace{-0.5cm}
\caption{\it \small A - $\Delta{u}^{c}_{NL}$ vs. $\tau$ for $u_r\simeq0.0015$ and $\eta\in[0.00015\,\textrm{(blue)},\,0.0025\,\textrm{(red)}]$. B - Mode evolution of the $3$ resonances (colored lines) and the respective single mode simulations (gray lines) for $\eta=0.00055$. C - Three resonance system ($\eta=0.00055$): bullets are $u_{r1}$ (green), $u_{r2}$ (yellow), $u_{r3}$ (blue), and $u_r\pm\Delta u_{NL}$ (solid lines), $u_r\pm\Delta u^{c}_{NL}$ (dashed lines) are denoted with corresponding colors.
\label{fig_nlvs}}
\end{figure}

\paragraph{Non-linear velocity spread and resonance overlap at saturation} The dynamics of one isolated wave consists of an initial exponential growth of the mode followed by the non-linear saturation, where the particles are trapped and begin to slosh back and forth in the potential well of the wave. This makes the mode intensity oscillate and generates rotating clumps in the phase-space. Assuming a single mode, the approximation of the post-saturation dynamics (we denote with $|\bar{\phi}|^{S}$ the mode saturation level) by an instantaneous harmonic oscillator allows to identify the so called trapping (bounce) frequency $\omega_B$ as $\bar{\omega}_B=\sqrt{2\ell^{2}|\bar{\phi}|^{S}}=\sqrt{2\beta}\;\ell\,\bar{\gamma}_L$ (where we assumed the quadratic scaling $|\bar{\phi}|^{S}=\beta\bar{\gamma}_L^{2}$ with $\beta=const.$). In order to characterize the non-linear dynamics, we introduce the clump width $\Delta{u}^{c}_{NL}$ defined by measuring the largest instantaneous velocity of particles initialized with $u<u_r$ and the smallest velocity of particles with $u<u_r$ at $\tau=0$. This measure is performed during the temporal evolution (see \figref{fig_nlvs}A, for a specific case) and $\Delta{u}^{c}_{NL}$ is taken as the value at saturation time $\tau_S$.
\vspace{-0.2cm}
We analyze 5 distinct cases having equispaced resonant velocities in $[0.0013,\,0.0017]$. For each case, 10 simulations equispacing $\eta$ in $[0.00015,\,0.0025]$ are studied (giving distinct drives $\bar{\gamma}_L$). We find that the trapping frequency behaves as $\bar{\omega}_B=(3.31\pm0.06)\;\bar{\gamma}_L$ (in agreement with well-known results in the literature \cite{ZCrmp}(and refs. therein)), while $\Delta u^{c}_{NL}/u_r=(6.64\pm0.12)\;\bar{\gamma}_L$.

Let us now set a system with $3$ modes and study the resonance overlap focusing only on the saturation time. We consider that resonance overlap takes place present when phase-space regions associated to different resonances can mix; and, for the analyzed case (namely $u_{r1}\simeq0.0013$, $u_{r2}\simeq0.0015$, $u_{r3}=0.0017$), the onset of the overlap regime occurs for $\eta\geqslant0.00055$ (see the mode evolution in \figref{fig_nlvs}B) due to the progressive enhancement of the non-linear velocity spread. If we depict the resonance position and the corresponding $\Delta u^{c}_{NL}$ (\figref{fig_nlvs}C, dashed lines) for the threshold value $\eta=0.00055$, the non-linear trapping regions appear, actually, non-overlapped. This evidence has the important physical implication that also particles which are not trapped by the wave near resonance \cite{EEbook} are relevant in the ``active'' overlap of different non-linear fluctuations. There is vast literature on this issues, related with the analysis of the transition to stochasticity of adjacent resonances, leading to the well-known Chirikov overlap criterion (for details, see \citer{LL10}). In the present case, the perturbations are not imposed externally as it is usually assumed \cite{EEbook,LL10}, but are self-consistently determined including fluctuation spectrum as well as beam particle nonlinear evolution. Similar to well-known literature \cite{LL10}, we find that a scale factor should be applied to the quantity $\Delta u^{c}_{NL}$ in order to obtain the overlap of the resonance width at saturation time, namely $\Delta u_{NL}\simeq \alpha \Delta u^{c}_{NL}$ with $\alpha\simeq1.28$ (\figref{fig_nlvs}C, solid lines). We stress how the resonance $u_{r1}$ (green curve) is initially isolated, but the synergistic
\begin{wrapfigure}{r}{0.4\textwidth}\vspace{-0.2cm}
\centering
\includegraphics[width=0.4\columnwidth]{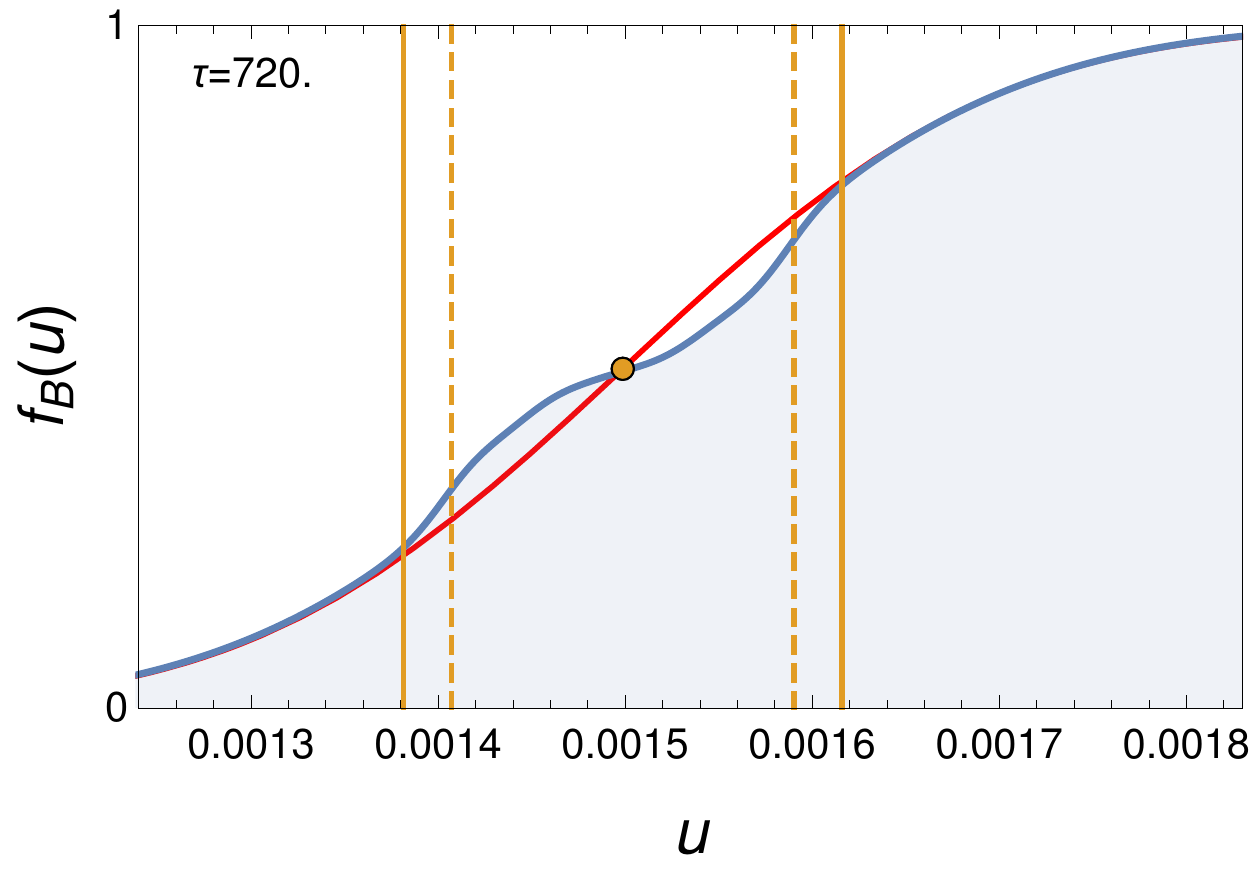}\vspace{-0.5cm}
\caption{\it \small Snapshot of $f_B(u,\tau_S=720)$ for $\eta=0.00055$ in the presence the single resonance $u_{r2}$ (bullet). The red line is $F_0$. We also indicate $u_r\pm\Delta u_{NL}$ (solid lines) and $u_r\pm\Delta u^{c}_{NL}$ (dashed lines).\label{fig_nlnlnl}}
\end{wrapfigure}
non-linear interaction of the two overlapping resonances modifies the dynamics and broadens the region affected by significant non-linear velocity spread. Thus, overlap with $u_{r1}$ sets in at later times, after fluctuations 2 and 3 have been nonlinearly amplified.

To shed light on the dynamic role of un-trapped particles, let now analyze the example of the function $f_B$ evolved until saturation in the presence of only the resonance $u_{r2}$ (\figref{fig_nlnlnl}): we remind that $\Delta u^{c}_{NL}/u_r\simeq6.64\;\bar{\gamma}_L$ and $\Delta u_{NL}/u_r\simeq8.5\;\bar{\gamma}_L$. It is evident how only the effective non-linear velocity spread $\Delta u_{NL}$ well characterizes the global distortion of $f_B(\tau_S)$ from $F_0$, accounting for un-trapped but nearly resonant particles and including effects at the plateau edges. This confirm the physical implication that also particles, which are not trapped by the wave near resonance, are relevant in the active overlap since the power transfer also involves those particles simultaneously feeling two (or multiple) modes.

In summary, the present analysis of the fully self-consistent evolution of the beam-plasma system agrees qualitatively and quantitatively with existing studies of the transition to the stochasticity made by dynamical system theory. Self-consistent non-linear effects seem to enhance $\Delta u_{NL}$ in the case of a broader spectrum (see \figref{fig_nlvs}C). We also emphasize that the present estimate is taken assuming the clump width at saturation. Different estimates could be obtained at later times and/or extending the time scale for the mixing of the phase space to occur.

\tiny NC would like to thank Ph. Lauber and T. Hayward for their fruitful discussions and advices. This work has been carried out within the framework of the EUROfusion Consortium [Enabling Research Projects: NLED (AWP15-ENR-01/ENEA-03), NAT (AWP17-ENR-MFE-MPG-01)] and has received funding from the Euratom research and training programme 2014-2018 under grant agreement No 633053. The views and opinions expressed herein do not necessarily reflect those of the European Commission.


\begin{thebibliography}{10}

\bibitem{CFMZJPP}
N.Carlevaro et al., \emph{J. Plasma Phys.} \textbf{81}, 495810515 (2015).

\bibitem{ncentropy}
N. Carlevaro et al., \emph{Entropy} \textbf{18}, 143 (2016).

\bibitem{malik16}
M. Idouakass et al., \emph{Phys. Plasmas} \textbf{23}, 102113 (2016).

\bibitem{ZCrmp}
L. Chen, F. Zonca, \emph{Rev. Mod. Phys.} \textbf{88}, 015008 (2016).

\bibitem{EEbook}
Y. Elskens, D.F. Escande, \emph{Microscopic Dynamics of Plasmas Chaos} (Taylor Francis Ltd) (2003).

\bibitem{LL10}
A.J. Lichtenberg, M.A. Lieberman, \emph{Regular and Chaotic Dynamics (2nd Edition)} (Springer-Verlag) (2010).

\end{thebibliography}
\end{document}